# Kinematic Total-Body 3D PET Video Reconstruction with 33-ms Frames via Hierarchical List-Mode EM and Validation-Adaptive Space-Time Regularisation

Andrew J. Reader, Gary J. R. Cook, Paul K. Marsden, Sarah-May Gould, Alexander Hammers and Simon R. Cherry

***Abstract*— Kinematic PET is introduced as an imaging task focused on visualisation of early radiotracer delivery and rapid multi-organ motion without gating. The proposed kinematic video reconstruction method uses 33-ms frames (~30 frames per second), initialised with 3D expectation maximisation (EM) time of flight (TOF) list-mode reconstruction, followed by cycles of a 2-stage update. These 2 stages are i) a hierarchical EM (HEM) update using multiple temporal pooling levels, and ii) a validation-adaptive space-time (VAST) regularisation update. HEM combines frame-specific EM corrections with corrections estimated from higher-count temporal blocks, slowing noise amplification for the ultrashort frames. VAST mixes each 3D target patch in a frame with source patches sampled nearby in space and time. It selects the candidate that best matches the mean of the corresponding patches in the two adjacent frames, with the unchanged patch also eligible. Optional temporal block sum correction uses the initial EM temporal sum as a reference to reduce loss of spatial detail. HEM+VAST was evaluated with a dynamic digital phantom, and demonstrated on human FDG data acquired on a Siemens Biograph Vision Quadra (28 - 58 s post-injection). At a background spatial coefficient of variation of 10%, the temporal full width at half maximum of the fastest phantom bolus was 151 ms with HEM+VAST, compared with 86 ms for the ground truth. The video frame rate FDG images reveal vascular and organ detail, radiotracer bolus passage and respiration-related multi-organ motion. Kinematic PET could complement conventional dynamic PET by supporting the development of image-derived physiological biomarkers and motion estimation for quantitative reconstruction. Future work should compare HEM+VAST with prior 4D PET reconstruction methods at the same 33-ms frame duration and matched noise levels.**

## I. INTRODUCTION

TOTAL-BODY positron emission tomography (TB PET) enables simultaneous molecular imaging over a large proportion, or the totality, of the human body. By extending the axial field of view (FOV), TB PET typically achieves 5-10 times greater sensitivity than conventional PET, offering lower-dose imaging, shorter acquisitions, and temporal framing that was previously impractical [1]. Nonetheless, PET remains count limited, with image reconstruction governed by trade-offs between temporal resolution, spatial resolution, and noise. When TB PET list-mode data are divided into frames of the order of tens of milliseconds, each 3D frame may contain only as few as hundreds of thousands of events. Such limited data, relative to the several million image voxels per volume, yields extremely noisy and sparse images.

Conventional dynamic PET uses frames lasting seconds to minutes to estimate regional or voxel-level time–activity curves (TACs) and kinetic parameters. The focus is on binding, metabolism, and clearance. However, at much higher frame rates we define a distinct and complementary imaging task: *kinematic PET*, where the goal is to operate at video frame rates for low-noise visualisation and assessment of early radiotracer bolus passage, rapid multi-organ displacements and cardiac contractions. Hence the kinematic imaging task does not necessitate quantitative image reconstruction, having the option to obviate the complications of accurate attenuation, scatter and randoms modelling for ultrashort frames. Instead, with its prioritisation of deformations, moving boundaries and transient features, kinematic PET can focus on spatiotemporal resolution and contrast at low noise.

Early progress towards kinematic PET, although not operating at true video frame rates, was demonstrated by Zhang et al. [2], using the uEXPLORER scanner. Total-body kernel expectation-maximisation (TB-KEM) was used to reconstruct 100-ms frames containing fewer than one million true coincidences. KEM represents each frame by spatial basis functions derived from higher-count image features [3],[4].

This work involved human subjects in its research. Approval of all ethical and experimental procedures and protocols was granted by the PET Research Review Committee through the TOTEM study (Total Body PET - Exploring new iMaging protocols for clinical research) under REC reference 24/LO/0878 (IRAS ID: 339077).

Andrew J. Reader and Paul K. Marsden are with the School of Biomedical Engineering and Imaging Sciences, King's College London, WC2R 2LS London, U.K. (e-mail: andrew.reader@kcl.ac.uk).

Gary J. R. Cook and Alexander Hammers are with the School of Biomedical Engineering and Imaging Sciences, King's College London, WC2R 2LS London, U.K., and with the Guy's and St Thomas' PET Centre, SE1 7EH London, U.K. (e-mail: gary.cook@kcl.ac.uk).

Sarah-May Gould is with the Guy's and St Thomas' PET Centre, SE1 7EH London, U.K. (e-mail: sarah-may.gould@kcl.ac.uk).

Simon R. Cherry is with the School of Biomedical Engineering and Imaging Sciences, King's College London, WC2R 2LS London, U.K., and with the Department of Biomedical Engineering, University of California, Davis (e-mail: simon.cherry@kcl.ac.uk).

Videos of both MIPs and slices of a 3D [$^{18}$F]FDG kinematic video reconstruction can be found here: https://youtu.be/guVoguyVH10 (24 cycles of HEM+VAST+BSC) and https://youtu.be/PPTq48LvAuE (comparing 12 cycles of HEM+VAST to spatial and spatiotemporally smoothed MLEM).

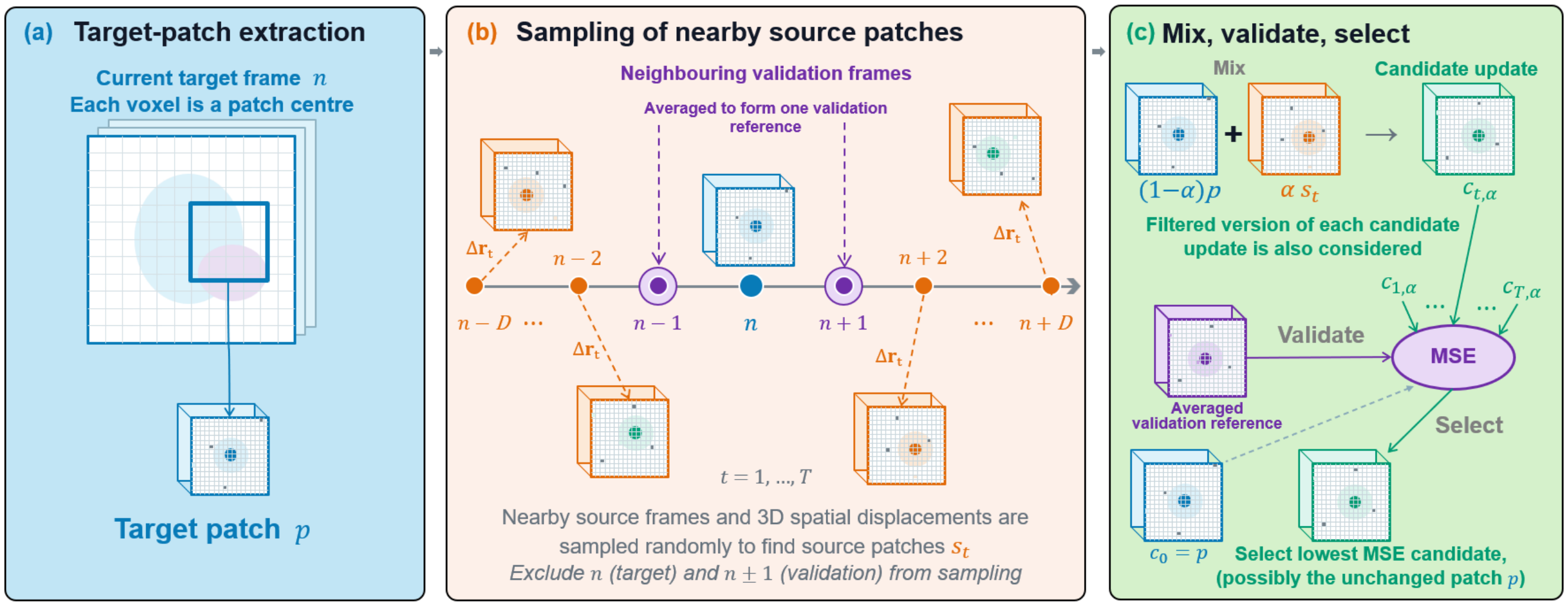

**Fig. 1.** VAST stage of the kinematic video reconstruction. For each 3D target patch, source patches are sampled from nearby non-validation frames, mixed with the target patch, to give candidate updates which are also optionally filtered and optionally sum matched. This collection of candidate updates (including the unchanged target patch) is each scored against the mean of the immediately time adjacent frames. Overlapping winning patches are averaged to form the output.

The feature vectors are derived from composite time–activity images and the resulting spatial kernel matrix is fixed during reconstruction. While kernel coefficients are estimated independently for each frame of PET data, the spatial basis functions learned from the composite images are applied for all frames. Hence moving boundaries, deformations, and transient features (which are poorly captured in composite images) are often inadequately represented.

Highly constrained backprojection (HYPR) also uses a higher-count composite image to provide high-quality spatial information to individual dynamic frames [5]. In particular, HYPR4D incorporates a data-driven 4D denoising operator within iterative reconstruction [6]. Non-local means and related filters estimate voxels from image neighborhoods judged to be similar [7]. Block-matching methods extend this principle: kinetics-based block matching with 5D filtering groups patches according to their time–activity behavior before transform-domain filtering [8], whereas guided block matching with 4D filtering uses projection data to guide matching before thresholding and Wiener-filtering [9]. While these methods can preserve structure, at extremely low counts the matching itself can become dominated by noise, and only relying on target-source similarity without an independent reference does not ensure improved estimation of the dynamic images. Maximum likelihood estimation of motion from short frames has been another route, but only so far demonstrated as viable at relatively long 0.5 second frames, too long to capture fast movement [10].

Deep learning provides further options for spatiotemporal regularisation. 4D deep-image-prior (DIP) approaches optimise a subject-specific network without paired low- and high-count training data [11], while population-based DIP methods additionally use information learned from multiple subjects [12]. Self-supervised denoising has been reported for noise-varying dynamic cardiac PET [13], and single-scan adaptive spatiotemporal graph filtering progressively estimates spatial and temporal correlations from a single acquisition [14]. These approaches usually achieve regularisation from the inductive prior of the network architecture, population training, or anatomical information. However none of these approaches have tackled reconstruction of ~1000 frames of video-rate 3D PET images from measured data.

## *A. Hierarchical EM and VAST*

We propose hierarchical expectation-maximisation (HEM) and validation-adaptive space–time (VAST) regularisation to address the challenges of kinematic PET video reconstruction. Starting from independently reconstructed ultrashort frames, the EM stage calculates an initial EM update image for every frame. These EM update images are then pooled across varying duration temporal blocks and combined before application to the individual frame images. This provides higher-count updates of spatial components shared across nearby frames while retaining distinct ultrashort-frame images and their temporal variation. After the HEM update, VAST regularisation (Fig. 1) subsequently considers each 3D patch in each volume, and searches for source patches which are spatiotemporally nearby in order to combine the best match (assessed by neighbouring validation frames) with each 3D patch and reduce noise. To assess the best match, candidate updates of each patch are evaluated with a validation reference formed by averaging the immediately preceding and following frames. The unmodified 3D target patch is itself considered as a candidate alongside the other updates. Winning updates for all the overlapping patches are averaged to form each output frame. This VAST denoising approach seeks information from elsewhere in the spacetime sequence to help fill in the gaps in the extremely sparse 3D frames. Repeating the HEM and VAST updates combines measured list-mode data EM corrections with validation-controlled propagation of information through space and time. Over several iterations, useful image information, displaced far in space and time, can make its way across spacetime to help denoise a given region of the image. Here, VAST is used as an algorithmic regulariser within an alternating EM reconstruction rather than as a single explicit penalised-likelihood objective. Whilst it can be extended to a formal regularising penalty in a MAP-EM nested or similar ADMM reconstruction (but with more complex hyperparameterisation) such further developments are left for future work, with the present work prioritising the development of the two separate components: HEM and VAST. Our novel contributions are:

a) Proposal and formulation of kinematic PET video reconstruction as an imaging task focused on visual assessment of rapid organ motion and early radiotracer transport, not necessitating quantitative image reconstruction
b) Hierarchical EM that uses overlapping temporal blocks of varying duration to slow noise amplification in the data-driven updates for ultrashort frames
c) VAST regularisation via local spacetime patch sampling and validation-controlled mixing for principled propagation of relevant image information in spacetime
d) Demonstration of human TB PET at 33-ms video rate framing (30 volumes/s), showing radiotracer-bolus passage, ungated cardiac and respiratory motion, and respiration-associated multi-organ motion.

We therefore extend TB PET to video frame rate imaging, with motion as key information for visualisation and assessment. This opens the way for new physiological biomarkers and potential exploitation for motion correction assistance in conventional quantitative PET reconstruction.

## II. Kinematic PET video reconstruction

This section describes the two key components to the kinematic video reconstruction methodology: hierarchical EM and VAST.

### *A. Frame-Level and Hierarchical EM (HEM)*

Time-of-flight (TOF) PET list-mode data are regarded as a series of coincident events with ms-precision time tags embedded. Conventional frame-level list-mode EM (LM-EM) reconstruction, for iterative update of a current estimate $\boldsymbol{x}_n^{(k)}$ of a 3D image for frame $n$, is given by

$$\boldsymbol{u}_n^{(k)} = \frac{\boldsymbol{x}_n^{(k)}}{\boldsymbol{s}} \sum_{l \in D_n} \frac{\boldsymbol{a}_l}{\boldsymbol{a}_l^{\top} \boldsymbol{x}_n^{(k)} + b_l}, \tag{1}$$

where $k$ indexes the reconstruction iteration (or cycle), $\boldsymbol{s}$ is the sensitivity image and $\boldsymbol{A}$ is the system matrix, with only row-access needed in (1) for the list-mode event indices $l$ within the LM data ($D_n$) for frame $n$. In the present implementation for kinematic PET reconstruction, $b_l$ is zero, since accurate modelling of scatter and randoms is not essential for the task of kinematic PET. However, $b_l$ can be included if reliable ultrashort list-mode randoms and scatter models are available. Independent application of (1) at 33-ms framing yields extremely noisy and sparse frame data (see later Fig. 2).

The proposed HEM update first calculates the standard independent EM updates $\boldsymbol{u}_n^{(k)}$ for every image frame. These are calculated once and used multiple times to form frame-level multiplicative correction images at multiple temporal scales. For temporal block $B$ at temporal hierarchy level $\ell$, the temporally-pooled EM gain is defined as:

$$\boldsymbol{g}_{\ell B}^{(k)} = \frac{\sum_{n \in B} w_{\ell B n} \, \boldsymbol{u}_n^{(k)}}{\sum_{n \in B} w_{\ell B n} \, \boldsymbol{x}_n^{(k)} + \epsilon \boldsymbol{1}}, \tag{2}$$

where $w_{\ell B n}$ are non-negative weights for each temporal level, designed and normalised such that their sum over the overlapping temporal blocks for each level equals one. The pooled ratio in equation (2) corresponds to weighted averaging of the frame-level EM correction images. While the correction images are shared temporally, they are applied separately to each image frame, and 4 different hierarchy block levels are considered in this present work (blocks of full duration, 4s, 2s and, importantly, the single frame level). Overlapping blocks are considered, with a stride length equal to half of the block length. Hence at each temporal block level $\ell$, the overlapping gains are combined to give a proposed image update for each frame $n$:

$$\boldsymbol{y}_{\ell n}^{(k)} = \boldsymbol{x}_n^{(k)} \sum_{B} w_{\ell B n} \, \boldsymbol{g}_{\ell B}^{(k)}. \tag{3}$$

For the pure single frame-level of the hierarchy, with no temporal pooling, the proposal is simply $\boldsymbol{y}_{\text{frame},n}^{(k)} = \boldsymbol{u}_n^{(k)}$. A major issue is of course noise, especially at the frame level, where if no list-mode data touches any given voxel, the outcome is a zero valued update, contributing significantly to the noise problem. Hence two policies are explored: retaining zeros (strict zeros) or using a neutral policy (where the gain value is simply kept at 1, leaving the current image value unmodified). As will be seen later, the neutral policy is more effective for noise reduction. The resulting proposals, $\tilde{\boldsymbol{y}}_{\ell n}^{(k)}$, after compensating for untouched voxels, are combined using non-negative hierarchy allocations $a_\ell$, with $\sum_\ell a_\ell = 1$. For positive current voxel values, the combined raw gain is

$$\boldsymbol{G}_n^{(k)} = \frac{\sum_\ell a_\ell \, \tilde{\boldsymbol{y}}_{\ell n}^{(k)}}{\boldsymbol{x}_n^{(k)}}. \tag{4}$$

Finally, one global relaxation $\lambda$ is then applied to push the gains towards the neutral gain case of one, followed by optional voxel-level clipping of this relaxed gain. The hierarchical LM-EM output is therefore:

$$\boldsymbol{z}_n^{(k)} = \boldsymbol{x}_n^{(k)} \, \text{clip}_{[g_{\min}, g_{\max}]} \left[1 + \lambda \, \boldsymbol{r}_n \big(\boldsymbol{G}_n^{(k)} - 1\big)\right], \tag{5}$$

with relaxation hyperparameter $\lambda$ ($0 < \lambda \le 1$) setting the overall data-consistency strength and $\boldsymbol{r}_n$ being an optional voxel-level sensitivity-adaptation factor. Hence the conventional EM frame updates are modified in four key ways, each seeking to mitigate the extreme noise and sparsity of the image frames: i) hierarchical temporal pooling, ii) neutral value handling for frame voxels receiving no data, iii) global EM relaxation and iv) clipping of the gain values. These pragmatic noise inhibiting adaptations do take the EM data consistency updates away from a pure ML framework, but they prove effective for non-quantitative kinematic PET video reconstruction.

### *B. Validation-Adaptive Spacetime Regularisation*

VAST (see again Fig. 1), operates on the current EM image sequence $\boldsymbol{z}_n^{(k)}$ from equation (5), whether obtained from frame-level EM or from HEM. For each frame $n$, the two adjacent frames are averaged to form a temporally separate image, regarded as a surrogate for a validation image:

$$\boldsymbol{v}_n^{(k)} = \frac{\boldsymbol{z}_{n-1}^{(k)} + \boldsymbol{z}_{n+1}^{(k)}}{2}. \quad (6)$$

Then, for every noisy 3D target patch $\boldsymbol{p}$ in a given frame, spatiotemporally nearby source patches are used as information to help fill in missing content. Since frames $n-1$, and $n+1$ are used for validation, these frames are excluded from consideration as potential source patches. A total of $T$ random sample patches are considered, taken from a limited range of temporally nearby source frames $m$, $2 \leq |m-n| \leq F_{max}$ each of which is randomly displaced according to isotropically-distributed radial displacements ($\leq R_{max}$). Trilinear interpolation of the displaced source frame can then supply a potential source patch $\boldsymbol{p}'$ to help denoise the target patch $\boldsymbol{p}$. A proposed candidate update for the target patch $\boldsymbol{p}$ is then formed by mixing the source and target:

$$\boldsymbol{c}_\alpha = (1-\alpha)\boldsymbol{p} + \alpha\boldsymbol{p}', \quad (7)$$

where the mixing coefficient $\alpha$ is found as the mean square error (MSE) minimum to get the candidate update to best match the validation patch $\boldsymbol{v}$:

$$\alpha^* = \frac{(\boldsymbol{v}-\boldsymbol{p})^{\mathrm{T}}(\boldsymbol{p}'-\boldsymbol{p})}{\| \boldsymbol{p}'-\boldsymbol{p} \|_2^2}. \quad (8)$$

Candidates with a negligible denominator or with $\alpha^* < \alpha_{\min}$ are rejected, else the candidate is evaluated with $\alpha = \min(\alpha^*, \alpha_{\max})$. The unmodified target patch $\boldsymbol{p}$ is kept as a null candidate. In addition to these candidate updates, VAST also considers filtered versions of the candidate patch updates, as well as sum matched versions. In the current implementation the filtering is a winsorisation (clamping) of the centre value within each $3^3$ neighbourhood within the candidate frame. If the centre voxel is the largest value, it is capped to match the second-largest value within the $3^3$ neighbourhood. This conservative filtering helps preserve spatial resolution and primarily targets isolated positive EM noise spikes. In addition, patch-sum-matched versions of the candidate patch updates are also considered as additional candidate update options:

$$\hat{\boldsymbol{c}}_\alpha = \boldsymbol{c}_\alpha \frac{\mathbf{1}^{\mathrm{T}}\boldsymbol{p}}{\mathbf{1}^{\mathrm{T}}\boldsymbol{c}_\alpha + \epsilon}, \quad (9)$$

where $\mathbf{1}^{\mathrm{T}}\boldsymbol{p}$ and $\mathbf{1}^{\mathrm{T}}\boldsymbol{c}_\alpha$ are simply patch sums. Every candidate update $\boldsymbol{c}$, including the unchanged patch $\boldsymbol{p}$, is stored in a list $C$, and each is assessed against the validation data $\boldsymbol{v}$ using the local validation MSE over the patch voxels:

$$\boldsymbol{c}^* = \arg\min_{\mathbf{c}\in C} \| \boldsymbol{c}-\boldsymbol{v} \|_2^2. \quad (10)$$

The winning candidate update of the patch, $\boldsymbol{c}^*$, contributes to the output. Overlapping winning patches are combined by normalised averaging, enabling us to define the VAST operator $\mathcal{V}$ for a frame $n$ of an image series $\boldsymbol{Z}$ as follows:

$$\mathcal{V}(\boldsymbol{Z})_n = \frac{\sum_{\mathrm{r}} \boldsymbol{P}_{\mathrm{r}}^{\mathrm{T}} \boldsymbol{c}_{n,\mathrm{r}}^*}{\sum_{\mathrm{r}} \boldsymbol{P}_{\mathrm{r}}^{\mathrm{T}} \mathbf{1} + \epsilon \mathbf{1}}, \quad (11)$$

where $\boldsymbol{P}_{\mathrm{r}}^{\mathrm{T}}$ places a patch back into the image grid and $\mathbf{c}_{n,\mathbf{r}}^*$ is the winner for frame $n$ and patch centre $\mathbf{r}$. Spatial boundaries use edge replication for scoring and displaced sampling, and the denominator in (11) accounts for reduced overlap at image edges. One completed pass through all frames in $\boldsymbol{Z}$ constitutes one VAST sweep. All target, source, and validation patches within a sweep are derived from the same previous temporal sequence $\boldsymbol{Z}^{(k)}$ $(= \boldsymbol{z}_1^{(k)} \dots \boldsymbol{z}_N^{(k)})$. The relaxed update from a VAST sweep is $\boldsymbol{Z} \leftarrow (1-\Lambda)\boldsymbol{Z} + \Lambda\widetilde{\boldsymbol{Z}}$, where $\widetilde{\boldsymbol{Z}}$ is the unrelaxed proposal. In the main configuration, $\Lambda = 0.75$.

*C. Alternating Reconstruction*

To give a warm start, 6 independent frame EM iterations are used to form the starting kinematic image series $\boldsymbol{X}^{(0)}$ $(=\boldsymbol{x}_1^{(0)} \dots \boldsymbol{x}_N^{(0)})$. Each subsequent cycle applies one HEM update stage $H_\lambda$ followed by the overall VAST operator $\mathcal{V}_\Lambda$, which can be one or more VAST sweeps (with 2 sweeps subsequently used in the main configuration in this work when VAST is interleaved with HEM):

$$\boldsymbol{Z}^{(k)} = H_\lambda(\boldsymbol{X}^{(k)}), \qquad \boldsymbol{X}^{(k+1)} = \mathcal{V}_\Lambda(\boldsymbol{Z}^{(k)}). \quad (12)$$

This approach is an alternating algorithmic regulariser. The net change in total activity can be monitored by the complete cycle change. A fixed point satisfies $\boldsymbol{X}^* = \mathcal{V}_\Lambda(H_\lambda(\boldsymbol{X}^*))$ (although the two stages may make opposing non-zero changes). The normalised complete-cycle change can be tracked by

$$\Delta_{\text{cycle}}^{(k)} = 100 \frac{\sum_{n,j}\left(X_{n,j}^{(k)} - X_{n,j}^{(k-1)}\right)}{\max\left\{\sum_{n,j}\left|X_{n,j}^{(k-1)}\right|, \epsilon\right\}}. \quad (13)$$

*D. Post-Processing: Temporal Block Sum Correction*

Optional final post-processing stages are proposed. The first is reapplication of winsorisation (clamping within successive 3D neighbourhoods of $3^3$ voxels), similar to the candidate update filtering used in VAST. As a post-processing option, this is implemented with sensitivity-image adaptation, allowing no filtering towards the centre of the FOV (where statistics are strongest), and the maximum level of filtering in regions of the image corresponding to low values of the sensitivity image $\boldsymbol{s}$. Here, for the regions of lowest sensitivity, winsorisation with clamping up to the 2nd highest outlier value was used. The second optional post-processing correction proposal is a *temporal block sum correction* (BSC). This is based on the observation that the overall temporal duration (e.g. 30 s) sum across frames of the initial noisy EM reconstruction (6 iterations in this work) can yield a high-quality image. When the HEM+VAST processed image is summed in a similar fashion, spatial resolution losses can be observed. So a temporal BSC, which can also be applied between updates (not done for the results presented here), allows correction for spatial resolution drift, by moving the voxel-level sums of all frames in the processed image towards those of the initial EM image. Specifically, in this work a full 30-second temporal BSC was considered, clipped to a max gain of 4 and a minimum gain of 0.25. At each voxel $j$, the initial EM series and the currently reconstructed/processed series were summed over the full-duration temporal block, giving $S_j^{(0)} = \sum_{n\in B} X_{n,j}^{(0)}$ and $S_j^{(pre)} = \sum_{n\in B} X_{n,j}^{(pre)}$ respectively. The reference-to-processed ratio was relaxed towards a value of 1 by $\gamma = 0.5$ before clipping, giving

$$g_j = \text{clip}_{[0.25,4]}\left(1 + \gamma\left[S_j^{(0)} / \left(S_j^{(pre)} + \epsilon\right) - 1\right]\right), \quad (14)$$

with $g_j$=1 when both sums were at or below $\epsilon$. The same voxel-specific gain was applied to every frame in the block, $X_{n,j}^{\mathrm{BSC}} = g_j X_{n,j}^{\mathrm{pre}}$, thereby rescaling each voxel's TAC without altering its shape or peak timing.

## III. Experimental Methods

### A. Simulated 3D TOF PET Data

The proposed kinematic PET video reconstruction methodology was evaluated using simulated data from a specially-designed TB-PET phantom. The phantom was a uniform cylindrical background with eight vessels (v1-v8) containing moving radiotracer boluses, a spatially uniform vessel with time-varying activity, an expanding and contracting lung (15 breaths/min), a centred spleen undergoing respiratory axial translations, and a beating heart (90 beats/min) with decaying activity. Three vessels had diameters of 6, 12 and 15 mm and bolus speeds of 40 mm/s. Five additional 12-mm vessels contained boluses moving at 35, 40, 60, 100 and 150 mm/s. The moving boluses had Gaussian axial profiles with a full width at half maximum (FWHM) of 10 mm (Fig. 2).

A total of ~41 million detected true coincidence events (8 s duration) were simulated from emission locations according to the time-dependent activity distribution of the digital phantom. At each location, a back-to-back photon pair was emitted uniformly over 4π and accepted only if both photons intersected a valid detector-crystal pair. Accepted coincidences were stored in a synthetic list-mode file with time-of-flight information and millisecond time tags. On average, over the 8 s, ~170k true coincidences per complete 33 ms frame were simulated (compared to ~300-500k prompt coincidences per frame for the subsequent real data reconstruction which covered 30 s).

Background noise in uniform regions of the cylinder was measured in each frame using the percentage coefficient of variation (CV, standard deviation divided by the mean), and summarised by the median across all time frames. Transverse and longitudinal widths were measured from radial vessel profiles and axial bolus profiles. Temporal width was estimated from the activity bolus time course at each vessel's midpoint. FWHM values were calculated using background-corrected, interpolated half-maximum crossings and expressed as percentages of the frame-matched ground-truth widths.

Truth-locked contrast-to-noise ratio (CNR) was calculated from the background-subtracted vessel signal sampled within ±5 mm of the known reference trajectory, divided by the background noise. Bolus speed was found using linear fitting to ridges in the $z$-position–time activity 2D maps. Heart, spleen and lung Dice scores were calculated from radial half-contrast boundaries and the median across frames was reported. Motion amplitude was estimated using sinusoidal fits to the outer-diameter oscillations of the heart and lung, and using axial centroid displacement for the spleen. CNR, width recovery, organ Dice and motion-amplitude recovery were compared for different reconstruction iterations/cycles and plotted with their background spatial CV to assess the trade off with image noise.

### B. Human Total-Body PET Data

Real TB PET data were obtained via the TOTEM study (Total Body PET - Exploring new iMaging protocols for clinical research; REC 24/LO/0878; IRAS 339077). An anonymised dynamic [$^{18}$F]FDG acquisition (214.51 MBq injected activity) was performed on a Siemens Biograph Vision Quadra PET-CT system. All of the prompt list-mode data were used in the reconstruction (full maximum ring difference 322, ~52° axial acceptance), for which NEMA sensitivity is 176 kcps/MBq [15]. The present work used 30 s of list-mode data acquired 28–58 s after injection, covering the time interval of early radiotracer delivery and organ uptake.

### C. Image Reconstruction and Implementation

The list-mode data were divided into 33-ms frames, giving 243 frames for the simulated case (8 s duration), and 910 frames for the real data (30 s). Images were reconstructed on a 200 × 200 × 350 grid with 3-mm isotropic voxels (600 × 600 × 1050 mm field of view). TOF LM-EM used a 35-mm-FWHM timing kernel, a geometrical sensitivity image, and image-space resolution modelling with a 5-mm-FWHM point-spread function [16], [17], [18]. Six independent frame EM iterations initialised the alternating reconstruction. Attenuation, scatter, randoms, detector-efficiency normalisation, and dead-time modelling were not included in this kinematic reconstruction

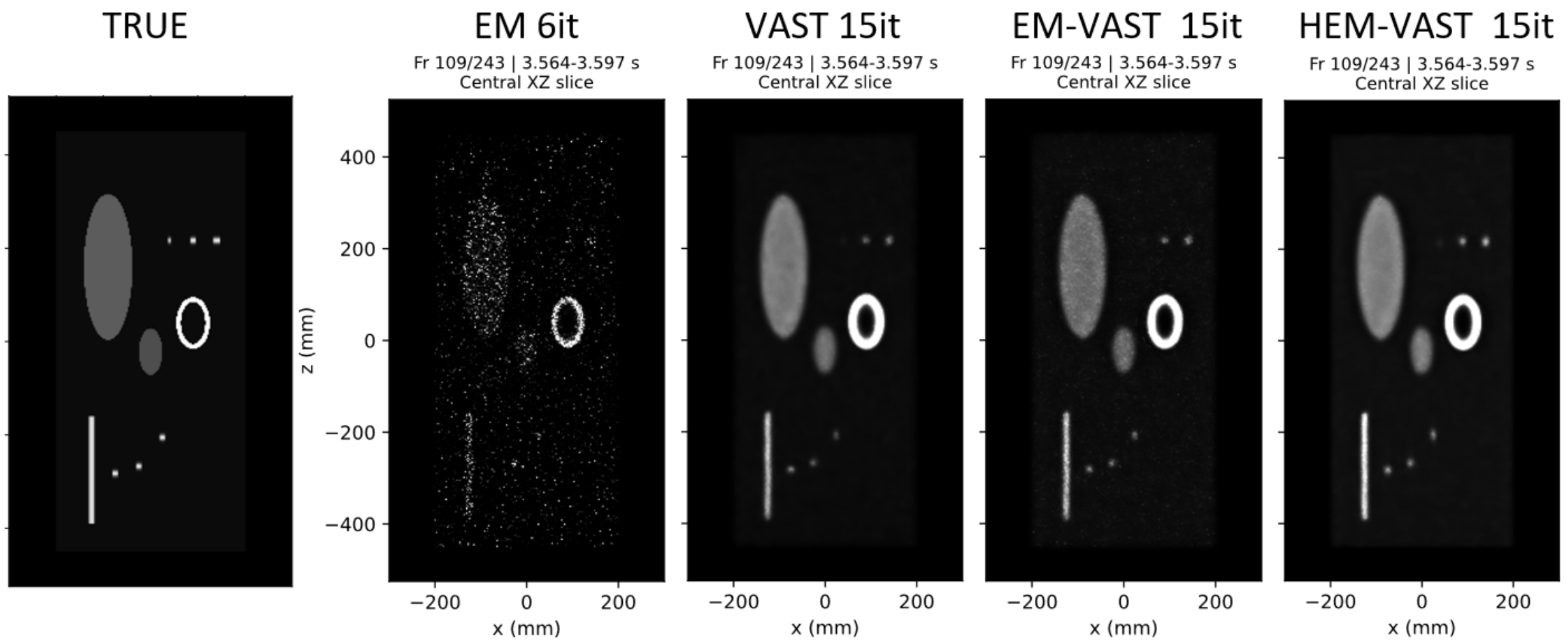

**Fig. 2.** Central coronal slices of a single 3D (200 × 200 × 350 voxels) 33-ms frame from the digital phantom for the 4 methods being compared: independent frame EM iteration 6, 15 iterations of VAST only applied after the 6 EM iterations, then for the case of interleaving the VAST updates with 10% relaxed frame EM, and finally with HEM also at a data consistency strength of 10%. To aid visibility, the 4 slices on the right are displayed at 20% of their maximum value, the reference ground truth image slice is displayed at 25% of its maximum value. These are single 2D slices from just 1 of 243 frames of 200×200×350 images. A video of the phantom is available, comparing HEM+VAST with smoothed MLEM: https://youtu.be/jioujNVapfM.

with its aforementioned focus on motion, resolution and contrast.

For the HEM stage, the 4 levels were the full duration of the reconstruction (30 s for real data, 8 s for simulated data), 4-s blocks, 2-s blocks and individual frames. The 4-s and 2-s blocks used 50% overlap. The relative weightings of the temporal levels (used in equation 4) were 20%, 20%, 20% and 40% for the frame level. The primary choice of HEM relaxation was $\lambda = 0.2$ and the clipping interval, applied after sensitivity-adaptive relaxation, was [0.5, 2]. VAST used $T$=32 trials per target frame and sweep, $F_{max}$ = 9 frames, $R_{max}$ = 8 mm, $11^3$-voxel patches centred at every voxel, α constrained to [0.05, 0.5], one-outlier $3^3$ upper winsorisation, matched and unmatched candidates, MSE validation, and final output relaxation of $\Lambda = 0.75$.

### D. Comparisons

Assessment of VAST for different search space sizes was explored, as well as frame-level EM vs. HEM for different $\lambda$.

## IV. Results

Example reconstructions of the phantom were shown in Fig. 2.

### A. Space-Time Search Extent

Fig. 3 shows 3 configurations of VAST: a very local search ($F_{max}$ = 3, $R_{max}$ = 2 mm) and a broad search ($F_{max}$ = 27, $R_{max}$ = 24 mm). The middle ground is represented by the red curve with diamonds, which is the reference (core) parameterisation from here on ($F_{max}$ = 9, $R_{max}$ = 8 mm). In terms of organ motion and spatiotemporal resolution, neither of the extreme search ranges (i.e. too local or too broad) perform well compared to the moderate range search, for the case of $T$=32 trials. Smoothing of conventional EM reconstruction reveals that spatial smoothing reduces noise at the cost of spatial resolution, and temporal smoothing improves CNR at the expense of motion amplitudes.

### B. Including Frame Level EM

Fig. 4 shows the impact of interleaving frame-level EM just before each VAST sweep. Whilst increasing the noise level, the main benefits of increased data consistency between VAST updates through frame level EM are improved spatial and temporal resolutions. Also, it is notable that using neutral gain updates for untouched voxels (voxels within a frame receiving no contribution at all – which is the vast majority of voxels at this data sparsity) provides lower noise than when zero updates are used. Hence, subsequent tests used neutral update handling for any zeros in the EM updates.

### C. Including Hierarchical EM

Fig. 5 shows the benefit in noise reduction of using the hierarchical EM update. The spatial and temporal resolution improvements of frame EM are broadly retained but at much reduced noise. The 20% data consistency method (H020n) is the one that was adopted for the subsequent real data reconstruction using HEM with VAST.

### D. Approximate Temporal Resolution Estimates

Based on the measured width ($W_{V8}$) of vessel v8 (true bolus velocity 150 mm/s), with a true calculated FWHM from the noise-free phantom of 86.05 ms, a temporal resolution estimate was made for each method ($\sqrt{W_{V8}^2 - (86.05)^2}$) at matched noise, Table I. The 20% data consistent HEM+VAST method offers the better temporal resolution estimate (124 ms) for a fixed spatial CV noise level of 10%.

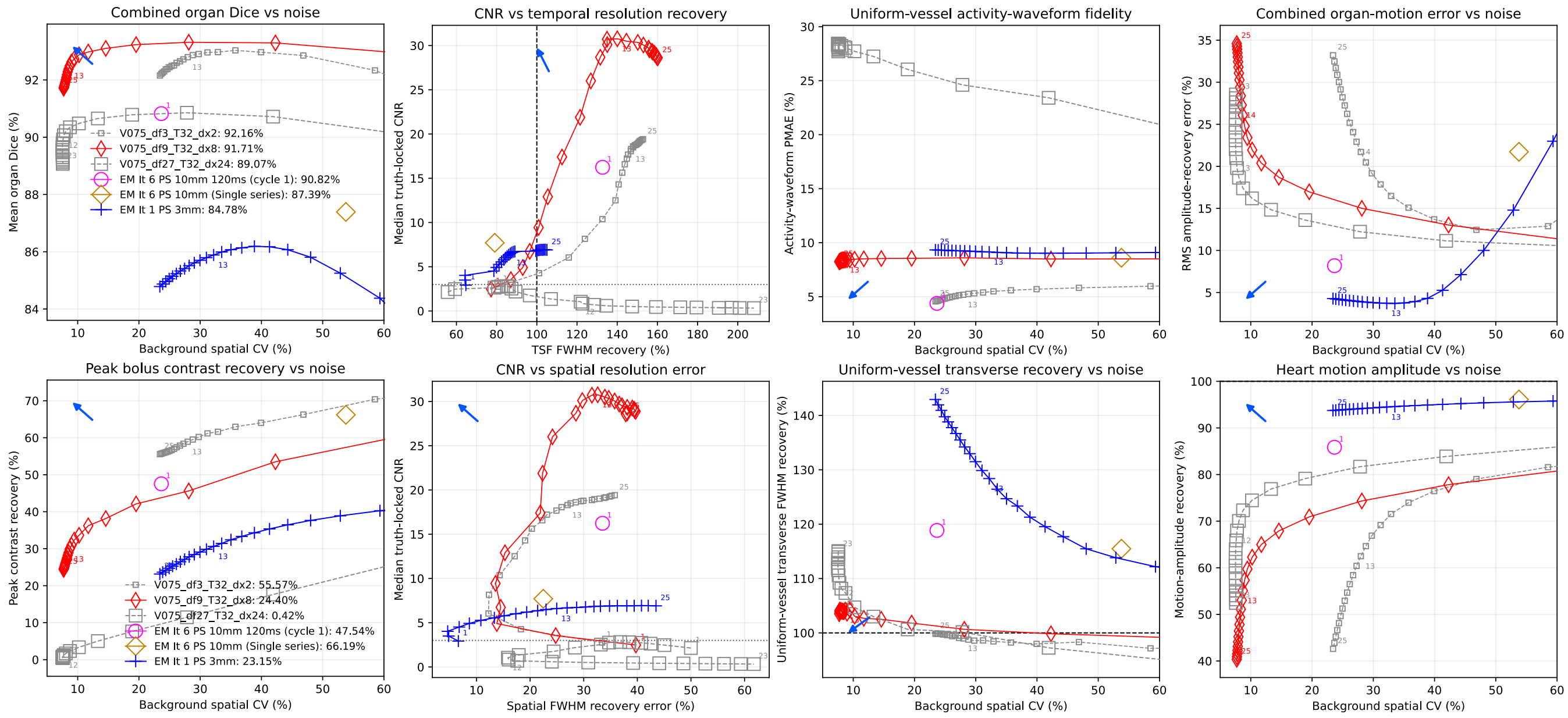

**Fig. 3.** VAST (V) performance compared to frame EM with post-smoothing (PS) (EM 1 iteration with 25 repeated 3 mm smooths, as well as EM 6 iterations with 10 mm PS, with and without 120 ms temporal smoothing). Top row: columns 1 to 4: 1) mean heart, lung and spleen Dice overlap versus background noise; 2) truth-locked CNR vs. temporal FWHM recovery; 3) normalised uniform-vessel activity-waveform percentage mean absolute error (PMAE) vs. noise; 4) combined organ-motion amplitude-recovery error vs. noise. Bottom row: 1) peak bolus contrast recovery vs. noise; 2) CNR vs. combined spatial FWHM recovery error; 3) uniform-vessel transverse FWHM recovery vs. noise; and 4) heart-motion amplitude recovery vs. noise. Combined organ-motion error is across all 3 organs, spatial resolution error is across longitudinal and transverse widths. Each point represents one selected iteration or cycle. Moving-vessel plots use medians across v1–v8. The blue arrow points in the direction of better performance. V075:VAST with $\Lambda$=0.75, (df=$F_{max}$ frames, dx= $R_{max}$ mm, $T$ trials).

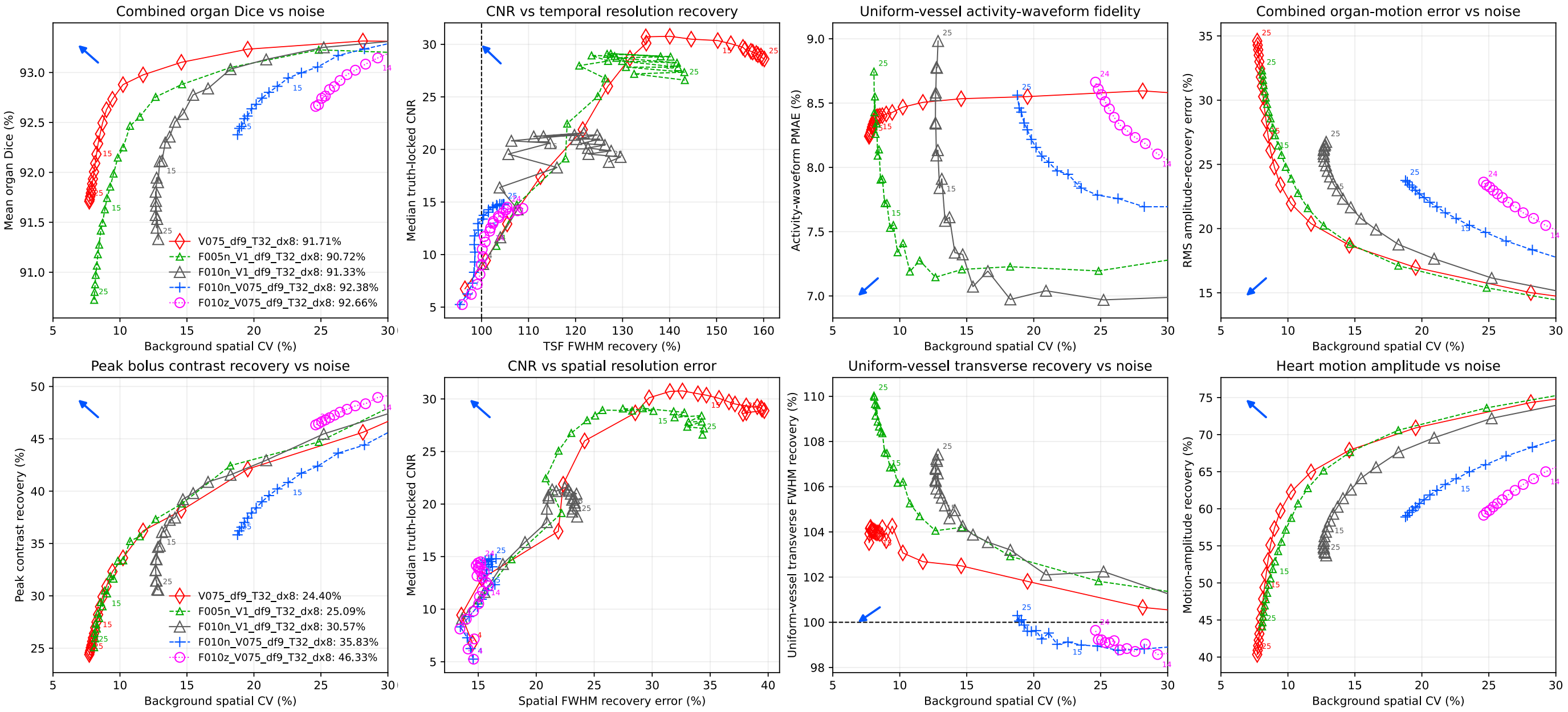

**Fig. 4.** Reconstruction performance for different data consistency strengths (5% (005) and 10% (010)) for when frame-level EM is used within the 2-stage cycle. Cycles/iterations 4 to 25 are shown for when frame EM is alternated with 1 VAST update with relaxation of either $\Lambda = 1$ (V1) or $\Lambda = 0.75$ (V075). The core parameters of the baseline VAST only processing are used in each case ($F_{max}$= 9 frames, $R_{max}$=8 mm and $T$=32 trials, (df9, dx8, T32)). For the 10% data consistency case, use of zeros (z) or a neutral update (n) for untouched voxels, is compared.

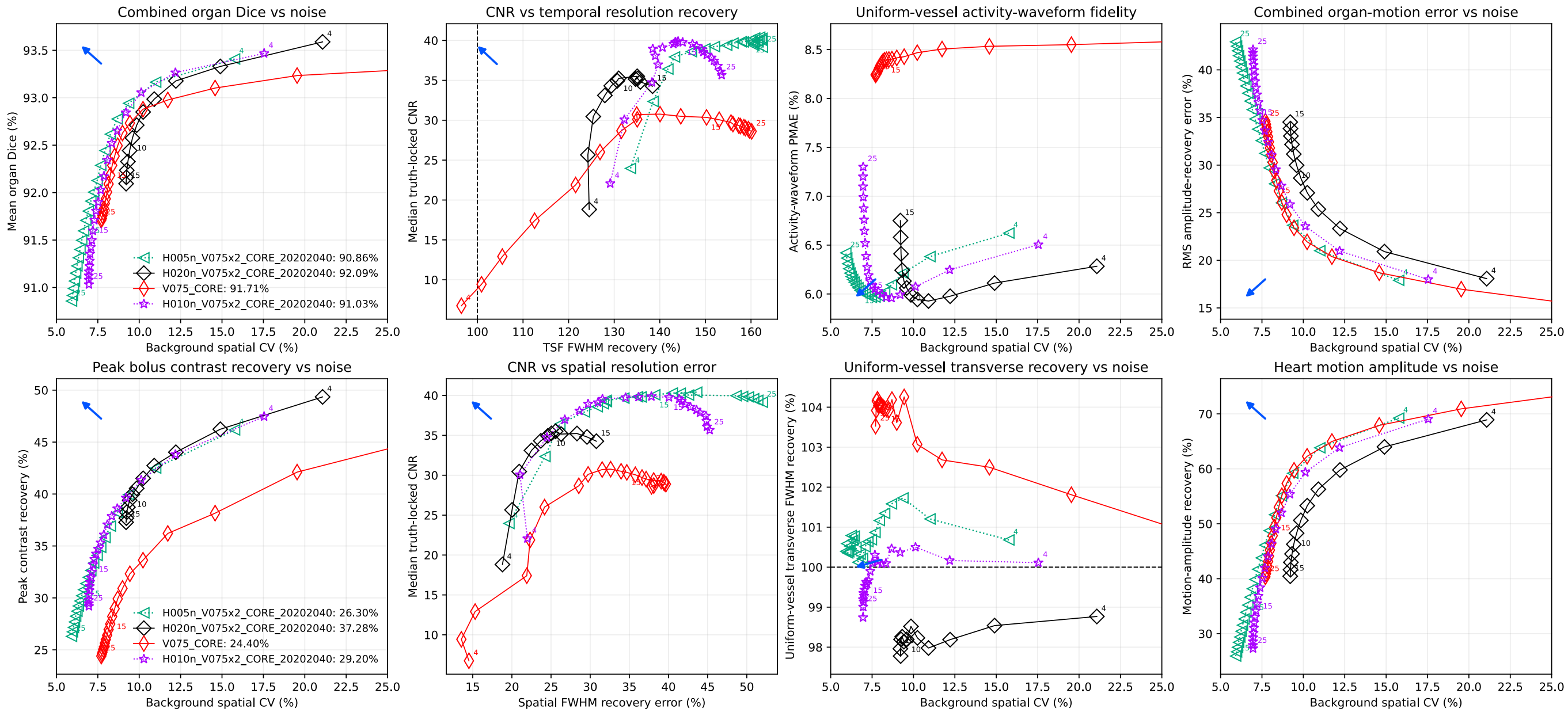

**Fig. 5.** Reconstruction performance comparison for different data consistency strengths (5%, 10% and 20%) for the HEM component of the 2-stage cycle. Cycles/iterations 4 to 25 are shown for when the HEM is interleaved with 2 VAST updates with 0.75 relaxation (V075×2). "CORE" refers to the core parameters of the baseline VAST only processing ($F_{max}$= 9 frames, $R_{max}$=8 mm and $T$=32 trials), and 20202040 refers to a 20/20/20/40% weighting scheme for the 4 levels.

TABLE I

TEMPORAL RESOLUTION ESTIMATES AT A NOISE LEVEL OF 10% SPATIAL CV

| Method | Cycles (equiv. cycle) | Median FWHM recovery | Median Bolus FWHM | V8 FWHM | Temporal-resolution estimate |
|---|---|---|---|---|---|
| **H020nV075×2** | 8–9 (8.61) | 130.2% | 324 ms | 151 ms | **124 ms** |
| **H010nV075×2** | 6–7 (6.13) | 138.5% | 330 ms | 159 ms | **134 ms** |
| **H005nV075×2** | 5–6 (5.61) | 140.5% | 332 ms | 165 ms | **141 ms** |
| **V075** | 10–11 (10.29) | 132.6% | 273 ms | 191 ms | **171 ms** |

### *E. Kinematic Video PET Reconstruction for Real Data*

Fig. 6 shows example slices and MIPs, and the linked video reveals passage of the [$^{18}$F]FDG bolus through the central vasculature, changing cardiac chamber activity, ungated cardiac contraction and respiration-associated movement of the spleen and kidneys. The supporting video is the main demonstration of the kinematic video reconstruction, with just static images in Fig. 6. Fig. 7 shows slices for a single frame from the proposed method compared to spatially and spatiotemporally smoothed MLEM. Supplemental Fig. 11 compares the same three methods after temporal BSC has been applied to each. Fig. 8 shows TACs for the methods at a boundary with the heart, revealing cardiac motion. This also reveals reduced amplitude of the cardiac-related TAC oscillation for the proposed method, although the TACs are not from images of matched background noise levels, with the EM methods not achieving the low noise level of the proposed approach. This comparison will in future work ideally need repeating for matched noise levels, by deriving robust noise estimates for real data. Fig. 9 shows fluctuations in a neck vessel TAC consistent with respiratory motion. Fig. 10 shows the level of change in the video reconstruction as a function of cycle during the two stages (EM and VAST) of each cycle.

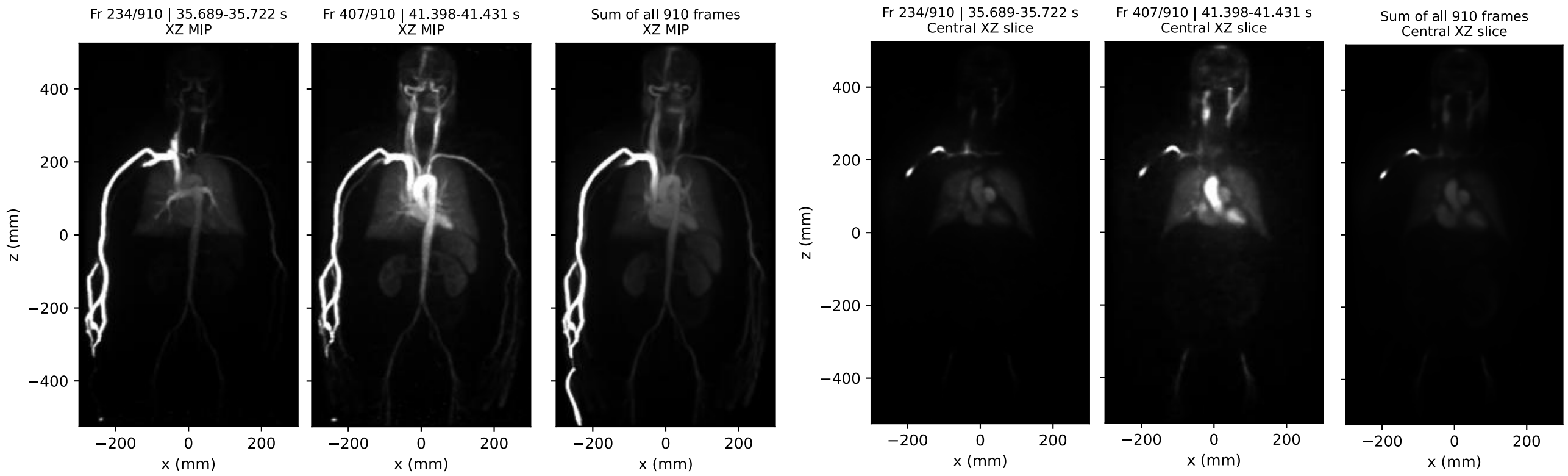

**Fig. 6.** Example real-data frames (left: MIPs, right: slices) for 16 cycles of the H020n_V075×2 method (HEM with 20% data consistency per stage, neutral updates for untouched voxels). No post-processing winsorisation, nor BSC, was applied to the above images. The sum of all 910 frames is also shown for each case. Video link for 24 cycles with BSC: [https://youtu.be/guVoguyVH10](https://youtu.be/guVoguyVH10).

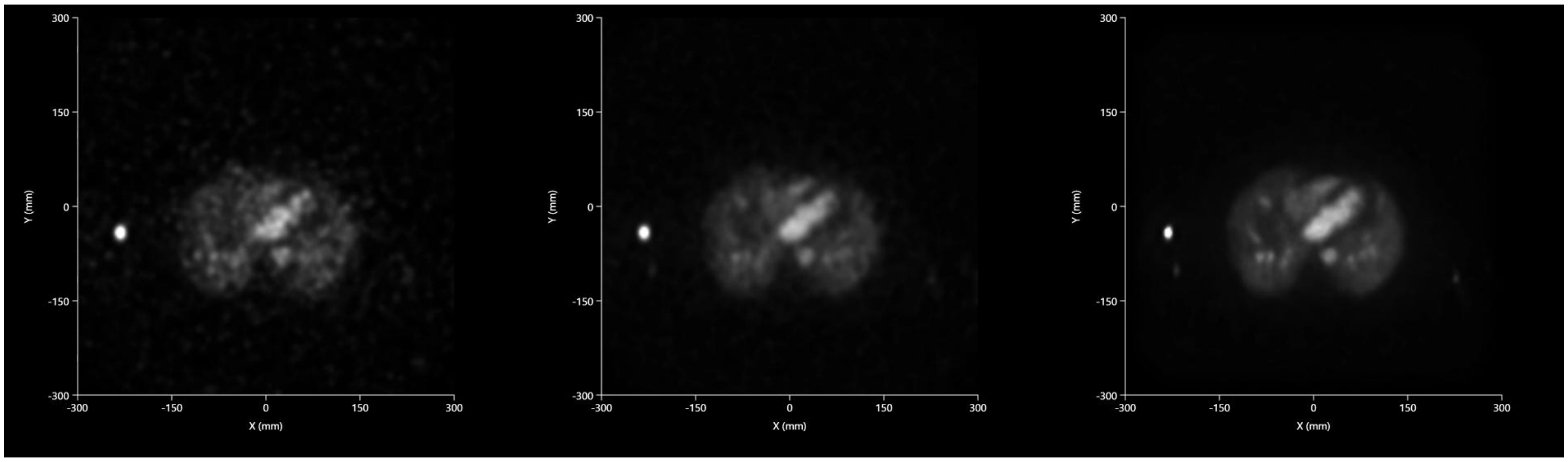

**Fig. 7.** Example real-data 33-ms frames (39.781-39.814 s) for a transverse plane (161) through the heart. Left: 6 iterations of MLEM with 10 mm spatial post smoothing. Middle: 6 iterations of MLEM with spatial post smoothing of 10 mm and temporal post smoothing of 120 ms. Right: 12 cycles of the proposed H020n_V075×2 method. No BSC was used for any of the images, but winsorisation was applied to H020n_V075×2 (of no consequence for the slice shown, as post-processing winsorisation targeted noise at the axial ends of the 3D image frames). Video comparison with and without BSC: [https://youtu.be/PPTq48LvAuE](https://youtu.be/PPTq48LvAuE).

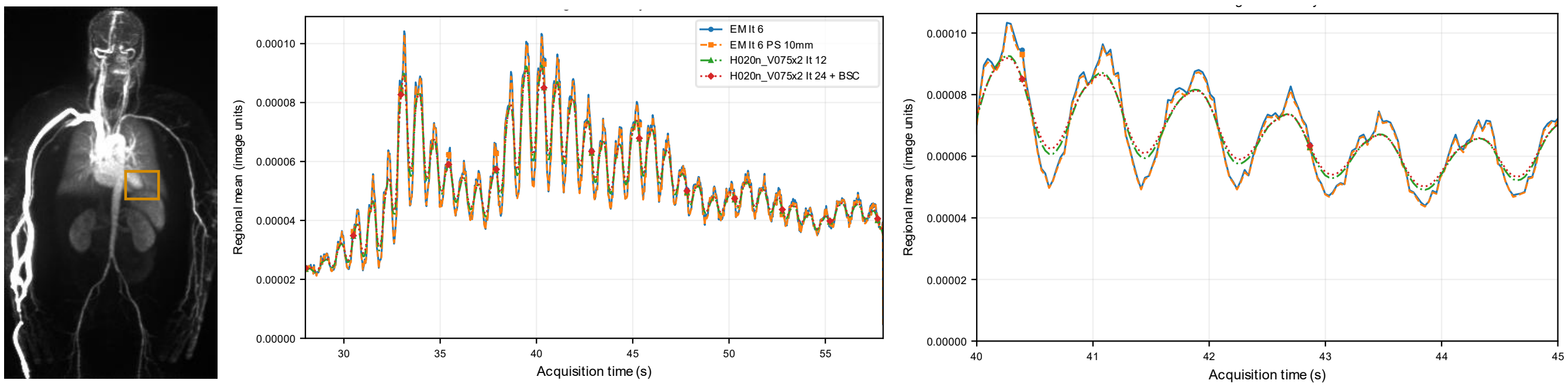

**Fig. 8.** TACs (28-58 s, and 40-45 s) for a volume (16926 voxels) over the heart (coronal MIP shown with the region on the summed EM iteration 6 reconstruction), related to cardiac motion. 13 oscillations are observed in the time interval 35 – 45 sec, corresponding to a heart rate of ~78 bpm.

### *F. Processing Times*

For 910 frames, using an NVIDIA RTX 5090 GPU and a 128 GB RAM workstation, a single VAST sweep took ~50-60 min, a single HEM step (with 4 temporal hierarchies) took up to 90 min, yielding ~3.5 hours per iteration for the selected parameterisation (H020n_V075×2). For the 243 frame simulation data, on an RTX 4090 GPU with a 64 GB workstation, the same methodology took ~50 min per iteration.

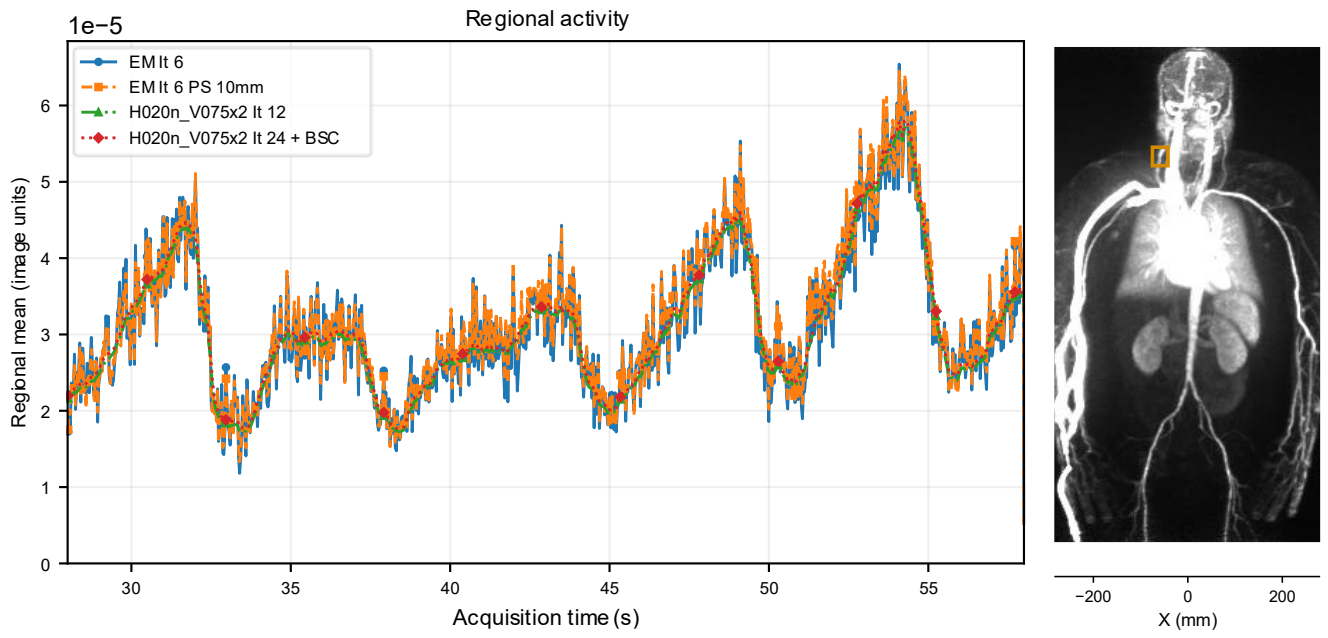

**Fig. 9**. TACs for a volume over a neck vessel (region shown in MIP on right).

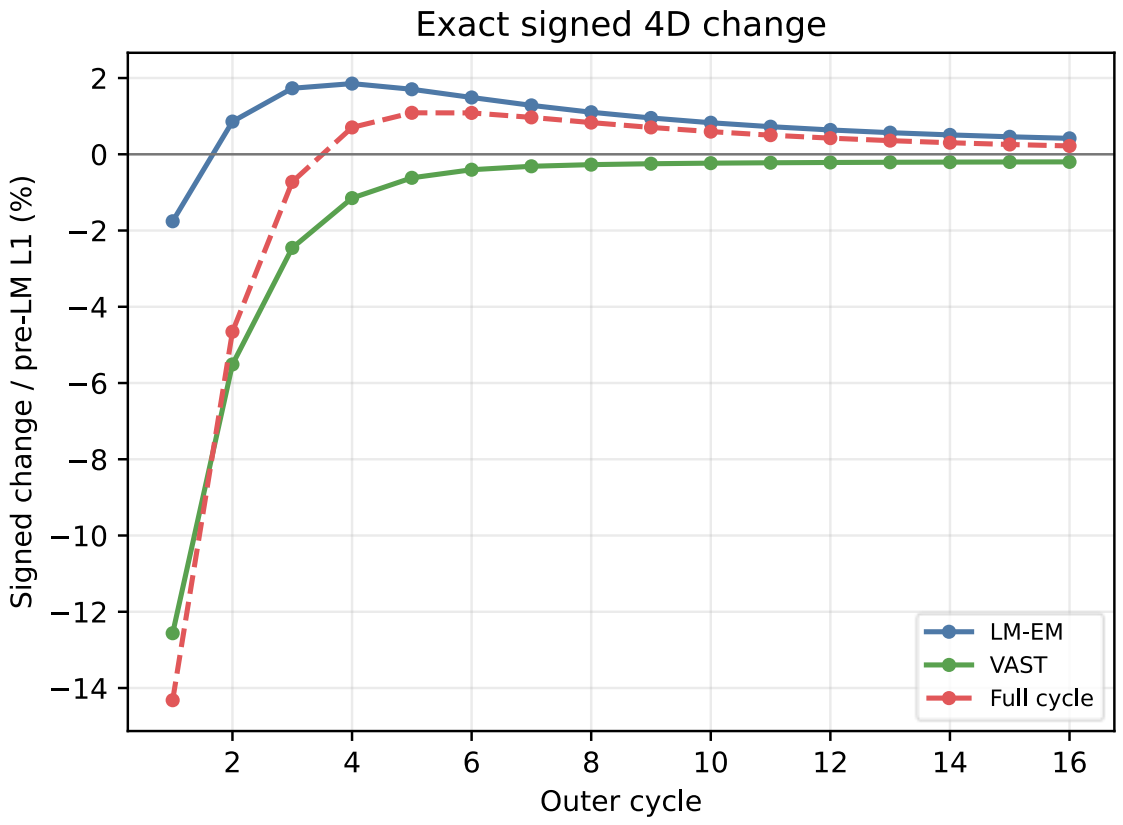

**Fig. 10**. Signed percentage change in total reconstructed activity over 16 cycles for real data (H020n_V075×2; 20% HEM data-consistency strength). Blue shows the change during hierarchical EM, green the combined change over two VAST sweeps, and dashed red their sum over the full cycle. Changes are normalised to the same pre-HEM image L1-norm for that cycle.

## V. Summary

Kinematic PET has been introduced as an imaging task focused on visualising early radiotracer delivery and multi-organ motion without gating, needing good spatiotemporal resolution and contrast at low noise. In proposing HEM+VAST, we have demonstrated one approach to this video reconstruction task by reconstructing TB PET data in 33-ms frames. An important next step is to compare HEM+VAST with a range of prior 4D PET reconstruction methods at the same frame duration and comparable image noise levels, assessing how well each method recovers rapid organ displacement, bolus contrast and the timing of tracer passage. Kinematic PET sequences could also support the development of image-derived physiological biomarkers of organ motion and radiotracer transport.

## Acknowledgment

The authors thank Yifan Zheng (Siemens Healthineers) for code to read Quadra list-mode data, and Sheut-Ling Lam for access to the anonymised [$^{18}$F]FDG list-mode data. Data were acquired via the King's College London and Guy's and St Thomas' PET Centre. The authors used ChatGPT and Codex (OpenAI) to partly assist writing of the Introduction section I, sections II and III. Real data were from an ethically approved study (TOTEM; REC 24/LO/0878; IRAS 339077). All authors declare that they have no known conflicts of interest in terms of competing financial interests or personal relationships that could have an influence or are relevant to the work reported in this paper. 

## References

[1] S. R. Cherry, R. D. Badawi, J. S. Karp, W. W. Moses, P. Price, and T. Jones, "Total-body imaging: Transforming the role of positron emission tomography," *Sci Transl Med,* vol. 9, no. 381, Mar 15 2017, doi: 10.1126/scitranslmed.aaf6169.

[2] X. Zhang, S. R. Cherry, Z. Xie, H. Shi, R. D. Badawi, and J. Qi, "Subsecond total-body imaging using ultrasensitive positron emission tomography," *Proc Natl Acad Sci U S A,* vol. 117, no. 5, pp. 2265–2267, Feb 4 2020, doi: 10.1073/pnas.1917379117.

[3] G. Wang and J. Qi, "PET image reconstruction using kernel method," *IEEE Trans Med Imaging,* vol. 34, no. 1, pp. 61–71, Jan 2015, doi: 10.1109/TMI.2014.2343916.

[4] G. Wang, "High Temporal-Resolution Dynamic PET Image Reconstruction Using a New Spatiotemporal Kernel Method," *IEEE Trans Med Imaging,* vol. 38, no. 3, pp. 664–674, Mar 2019, doi: 10.1109/TMI.2018.2869868.

[5] B. T. Christian, N. T. Vandehey, J. M. Floberg, and C. A. Mistretta, "Dynamic PET denoising with HYPR processing," *J Nucl Med,* vol. 51, no. 7, pp. 1147–54, Jul 2010, doi: 10.2967/jnumed.109.073999.

[6] C. W. J. Bevington, J.-C. Cheng, and V. Sossi, "A 4-D Iterative HYPR Denoising Operator Improves PET Image Quality," *IEEE Transactions on Radiation and Plasma Medical Sciences,* vol. 6, no. 6, pp. 641–655, 2022, doi: 10.1109/TRPMS.2021.3123537.

[7] J. Dutta, R. M. Leahy, and Q. Li, "Non-local means denoising of dynamic PET images," *PLoS One,* vol. 8, no. 12, p. e81390, 2013, doi: 10.1371/journal.pone.0081390.

[8] K. Ote *et al.*, "Kinetics-Induced Block Matching and 5-D Transform Domain Filtering for Dynamic PET Image Denoising," *IEEE Transactions on Radiation and Plasma Medical Sciences,* vol. 4, no. 6, pp. 720–728, 2020, doi: 10.1109/TRPMS.2020.3000221.

[9] L. Xin, W. Zhuo, H. Liu, and T. Xie, "Guided block matching and 4-D transform domain filter projection denoising method for dynamic PET image reconstruction," *EJNMMI Phys,* vol. 10, no. 1, p. 59, Sep 25 2023, doi: 10.1186/s40658-023-00580-5.

[10] R. Jose Santo *et al.*, "Correcting fast irregular motion in PET: maximum-likelihood motion and activity (MLMA) reconstruction," *EJNMMI Phys,* vol. 13, no. 1, May 26 2026, doi: 10.1186/s40658-026-00894-0.

[11] F. Hashimoto, H. Ohba, K. Ote, A. Kakimoto, H. Tsukada, and Y. Ouchi, "4D deep image prior: dynamic PET image denoising using an unsupervised four-dimensional branch convolutional neural network," *Phys Med Biol,* vol. 66, no. 1, p. 015006, Jan 14 2021, doi: 10.1088/1361-6560/abcd1a.

[12] Q. Liu *et al.*, "Population-based deep image prior for dynamic PET denoising: A data-driven approach to improve parametric quantification," *Med Image Anal,* vol. 95, p. 103180, Jul 2024, doi: 10.1016/j.media.2024.103180.

[13] H. Xie *et al.*, "Noise-aware dynamic image denoising and positron range correction for Rubidium-82 cardiac PET imaging via self-supervision," *Med Image Anal,* vol. 100, p. 103391, Feb 2025, doi: 10.1016/j.media.2024.103391.

[14] S. Guo, X. Li, S. Pan, and D. Zhang, "Single-scan adaptive graph filtering for dynamic PET denoising by exploring intrinsic spatio-temporal structure," *Front Med (Lausanne),* vol. 12, p. 1659122, 2025, doi: 10.3389/fmed.2025.1659122.

[15] G. A. Prenosil *et al.*, "Performance Characteristics of the Biograph Vision Quadra PET/CT System with a Long Axial Field of View Using the NEMA NU 2-2018 Standard," *J Nucl Med,* vol. 63, no. 3, pp. 476–484, Mar 2022, doi: 10.2967/jnumed.121.261972.

[16] A. J. Reader, K. Erlandsson, M. A. Flower, and R. J. Ott, "Fast accurate iterative reconstruction for low-statistics positron volume imaging," (in English), *Physics in Medicine and Biology,* vol. 43, no. 4, pp. 835–846, Apr 1998. doi: 10.1088/0031-9155/43/4/012.

[17] A. J. Reader *et al.*, "One-pass list-mode EM algorithm for high-resolution 3-D PET image reconstruction into large arrays," (in English), *IEEE Transactions on Nuclear Science,* vol. 49, no. 3, pp. 693–699, Jun 2002, doi: 10.1109/TNS.2002.1039550.

[18] A. J. Reader, P. J. Julyan, H. Williams, D. L. Hastings, and J. Zweit, "EM algorithm system modeling by image-space techniques for PET reconstruction," (in English), *IEEE Transactions on Nuclear Science,* vol. 50, no. 5, pp. 1392–1397, Oct 2003, doi: 10.1109/Tns.2003.817327.

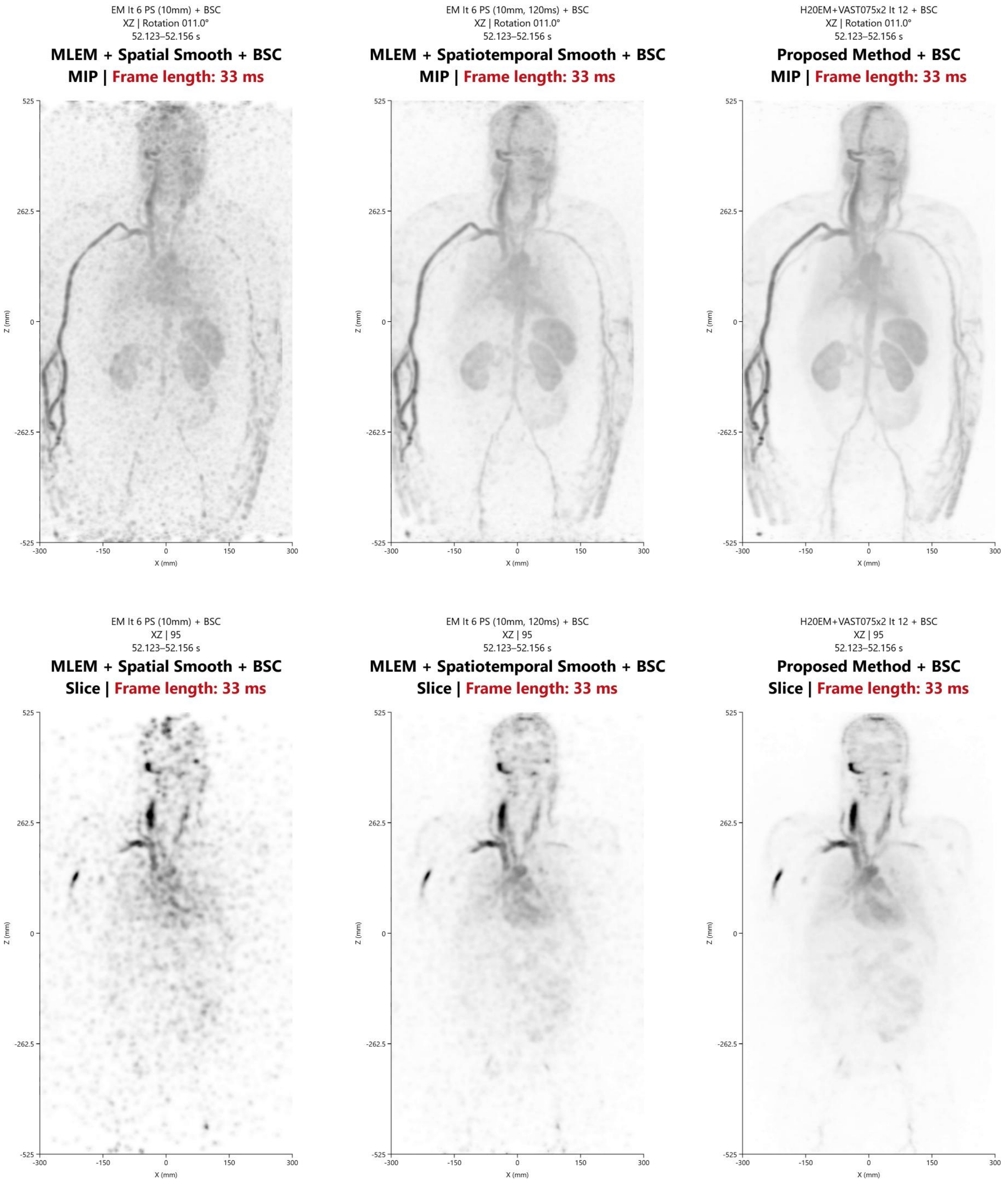

**Supplemental Fig. 11.** Example real-data frames from the video (https://youtu.be/PPTq48LvAuE), top: rotated MIPs, bottom: coronal slices, for 6 iterations of MLEM with 10 mm spatial post smoothing (left), 6 iterations of MLEM with spatial post smoothing of 10 mm and temporal post smoothing of 120 ms (middle) and 12 cycles of the proposed H020n_V075×2 method (right). In this case, BSC was applied as post-processing to all 3 methods, and winsorisation was applied to H020n_V075×2 prior to the temporal BSC.